\documentclass[12pt]{article}
\usepackage{amssymb,amsmath,bm}
\usepackage{epsf}
\usepackage{epsfig}
\usepackage{afterpage}
\usepackage{longtable}

\def\simge{\mathrel{\rlap{\raise 0.511ex \hbox{$>$}}{\lower 0.511ex \hbox{$\sim$}}}}
\def\simle{\mathrel{\rlap{\raise 0.511ex \hbox{$<$}}{\lower 0.511ex \hbox{$\sim$}}}}
\def\slash#1{\setbox0=\hbox{$#1$}\dimen0=\wd0
      \setbox1=\hbox{/} \dimen1=\wd1 \ifdim\dimen0>\dimen1
      \rlap{\hbox to \dimen0{\hfil/\hfil}} #1                        \else
      \rlap{\hbox to \dimen1{\hfil$#1$\hfil}}
      /   \fi}

\newcommand{\lsim}{
\mathrel{\hbox{\rlap{\hbox{\lower4pt\hbox{$\sim$}}}\hbox{$<$}}}}

\newcommand{\gsim}{
\mathrel{\hbox{\rlap{\hbox{\lower4pt\hbox{$\sim$}}}\hbox{$>$}}}}

\newcommand{\be}{\begin{equation}}
\newcommand{\ee}{\end{equation}}
\newcommand{\bea}{\begin{eqnarray}}
\newcommand{\eea}{\end{eqnarray}}

\newcommand{\bi}{\begin{itemize}}
\newcommand{\ei}{\end{itemize}}

\renewcommand{\baselinestretch}{1.2}
\newcommand{\newsection}[1]{\section{#1}\setcounter{equation}{0}}

\begin{document}

\begin{titlepage}
\vspace*{-0.5truecm}


\begin{flushright}
TUM-HEP-{647}/06
\end{flushright}

\vfill

\begin{center}
\boldmath

{\Large\textbf{Another Look at the Flavour Structure 
\vspace{0.3truecm}\\of
    the Littlest Higgs Model with T-Parity}}

\unboldmath
\end{center}

\vspace{0.4truecm}

\begin{center}
{\bf Monika Blanke, Andrzej J.~Buras, Anton Poschenrieder,\\
  Stefan Recksiegel,
 Cecilia  Tarantino, Selma Uhlig and Andreas Weiler
}
\vspace{0.4truecm}

 {\sl Physik Department, Technische Universit\"at M\"unchen,
D-85748 Garching, Germany}

\end{center}
\vspace{0.6cm}
\begin{abstract}
\vspace{0.2cm}
\noindent 
 We discuss the mixing matrix $V_{Hd}$ that describes the charged and 
 neutral current interactions between ordinary down-quarks and up- 
 and down-mirror quarks in the Littlest Higgs Model with T-parity (LHT).
 We point out that this matrix in addition to three mixing angles 
 contains {\it three} physical complex phases and not only one as used 
 in the present literature. We explain the reason for the presence of two
 additional phases, propose a new standard parameterization of $V_{Hd}$ and 
 briefly comment on the relevance of these new phases for the 
 phenomenology of FCNC processes in the LHT model.
In a separate paper we present a detailed numerical analysis, including these
 new phases, of $K$ and $B$ physics, with particular attention to the most
 interesting rare decays.
\end{abstract}\renewcommand{\baselinestretch}{1.2}

\vfill\vfill

\end{titlepage}

 \newsection{Introduction}
The Little Higgs models \cite{oldLH,LHreview,l2h} offer an alternative route
to the solution of the hierarchy problem. One of the most attractive models of
this class is the Littlest Higgs Model with T-parity (LHT) \cite{tparity}
which evades the stringent electroweak precision constraints Little
Higgs models usually have to cope with \cite{mH}.
In this model, the new gauge bosons, 
fermions and scalars are sufficiently light to be 
discovered at LHC and there is a dark matter candidate \cite{LHTphen}. Moreover, 
the flavour structure of the LHT model is richer than the one of the 
Standard Model (SM), mainly due to the presence of three doublets 
of mirror quarks and three doublets of mirror leptons and their 
weak interactions with the ordinary quarks and leptons.

As discussed first in \cite{Hubisz} and subsequently in \cite{BBPTUW}, 
the interactions of 
mirror quarks with ordinary quarks are described by two $3\times 3$ unitary 
mixing matrices $V_{Hd}$ and $V_{Hu}$ which satisfy 
\be
V_{Hu}^\dagger V_{Hd}=V_{\rm CKM}\,,
\label{eq:VHuVHd}
\ee
with $V_{\rm CKM}$ being the CKM matrix \cite{ckm}. Analogous matrices 
in the lepton sector exist. As emphasized in \cite{BBPTUW}, 
the presence of these 
new matrices implies new flavour and in particular CP-violating 
interactions that are absent in models with minimal flavour violation
(MFV) \cite{mfv1,mfv2,mfvearly}. 

These new interactions are mediated by heavy charged gauge bosons 
$W_H^\pm$ and neutral gauge bosons $Z_H$ and $A_H$ and, at higher
order, by the scalar triplet $\Phi$ with $V_{Hd}$ 
describing both the charged and neutral current interactions 
of standard down-quarks $d,s,b$ with mirror quarks. $V_{Hu}$
describes the corresponding interactions of standard up-quarks $u,c,t$. 

Present LHT analyses of particle-antiparticle 
mixing and CP-violation in $\Delta F=2$ processes, performed in \cite{Hubisz}
and in the first version of \cite{BBPTUW}, adopted for $V_{Hd}$ precisely the
standard parameterization of the CKM matrix in terms of three mixing angles
$\theta_{12}^d, \theta_{13}^d, \theta_{23}^d$ and one physical complex 
phase $\delta_{13}^d$.

In the present note we would like to point out that Hubisz et al. in
\cite{Hubisz}  and ourselves in the first version of \cite{BBPTUW} overlooked
the presence of two additional complex phases $\delta_{12}^d$ and 
$\delta_{23}^d$ in $V_{Hd}$ that, contrary to the CKM matrix, cannot be 
removed by phase transformations  and are physical. The new insight in 
the structure of $V_{Hd}$ was made in the context of a long and detailed 
analysis 
of rare $K$ and $B$  decays in the LHT model \cite{BBPRTUW}. 
As the issue in question 
has a more general character than the analysis in \cite{BBPRTUW}, 
it deserves in 
our opinion a separate note, that otherwise would get lost in a very 
long paper.

Below we demonstrate in simple terms the necessity for 
the presence of two new phases in 
$V_{Hd}$ and give a new parameterization of this matrix. 
The same claim will be then achieved from a more general method of counting
parameters.
We conclude 
with a few comments on the implications of our findings 
for FCNC processes.

\boldmath
\newsection{New Insight in the $V_{Hd}$ Matrix}
\unboldmath
\subsection{Simple Counting of Physical Phases}
In this section we will explicitly show that the parameterization of the
$V_{Hd}$ matrix requires not only one but three complex
phases, in addition to three mixing angles.

For simplicity we start considering the well-known CKM matrix $V_\text{CKM}$
\cite{ckm} that, due to unitarity, has in principle $3$ mixing angles and $6$ 
phases.
An $N \times N$ unitary matrix, in fact, is described by
$N(N-1)/2$ real parameters and $N(N+1)/2$ complex phases.
 Recalling that the CKM matrix appears in  charged, $W^\pm$ mediated,
 weak interactions between an up-quark and a down-quark, one has the
 additional freedom to eliminate some of the $V_\text{CKM}$ phases varying the
 phase of each quark state independently.
The number of phases that can be eliminated is $2N-1=5$, as
$V_\text{CKM}$ is
left invariant under an over-all phase change of all the quark fields.
This explains why the CKM matrix has $4$ independent parameters: $3$ mixing
angles and $1$ phase.

In the LHT Model, in addition to the SM flavour interactions described by
$V_\text{CKM}$, there are new interactions, mediated by the heavy gauge bosons
$W_H^\pm$, $Z_H$ and $A_H$, involving a SM and a mirror quark.
As discussed first in \cite{Hubisz} and subsequently in \cite{BBPTUW} 
the interactions of 
mirror quarks with ordinary quarks are described by two $3\times 3$ 
unitary mixing matrices $V_{Hd}$ and $V_{Hu}$, 
related through~(\ref{eq:VHuVHd}).
In the following discussion we will consider only $V_{Hd}$, while $V_{Hu}$ 
can be easily extracted from~(\ref{eq:VHuVHd}).

The mixing matrix $V_{Hd}$ is involved in the interactions of an ordinary
down-quark with either an up-mirror quark ($W_H^\pm$ mediated), or a
down-mirror quark ($Z_H$ or $A_H$ mediated).
From the unitarity of $V_{Hd}$ we know that it contains $3$ mixing angles
and $6$ complex phases.
Similarly to $V_\text{CKM}$, we can eliminate from $V_{Hd}$ some of the phases
by rotating the interacting states.
In this case, however, we have less freedom.
The phases of the standard fields, in fact, have been already chosen as to 
eliminate the maximum number of phases from $V_\text{CKM}$.
Acting on the mirror states only three phases can be still rotated away from 
$V_{Hd}$, which turns out to be
parameterized in terms of $3$ mixing angles and $3$ phases. 

We further note that once the phases of up-mirror quarks have been varied, the
same phase-rotation has to be applied to down-mirror quarks, since both
these fields are involved in the interaction described by $V_{Hd}$ with
ordinary down-quarks. This means that a phase-rotation of mirror quarks is
indeed able to eliminate only three phases from $V_{Hd}$.
An alternative and more general way of counting independent parameters, which 
confirms this result, will be provided in the next section.

The mixing matrix $V_{Hd}$ can be conveniently parameterized, generalizing
the usual CKM parameterization, as a product of three rotations, and
introducing a complex phase in each of them, thus obtaining
\be
\addtolength{\arraycolsep}{3pt}
V_{Hd}= \begin{pmatrix}
1 & 0 & 0\\
0 & c_{23}^d & s_{23}^d e^{- i\delta^d_{23}}\\
0 & -s_{23}^d e^{i\delta^d_{23}} & c_{23}^d\\
\end{pmatrix}\,\cdot
 \begin{pmatrix}
c_{13}^d & 0 & s_{13}^d e^{- i\delta^d_{13}}\\
0 & 1 & 0\\
-s_{13}^d e^{ i\delta^d_{13}} & 0 & c_{13}^d\\
\end{pmatrix}\,\cdot
 \begin{pmatrix}
c_{12}^d & s_{12}^d e^{- i\delta^d_{12}} & 0\\
-s_{12}^d e^{i\delta^d_{12}} & c_{12}^d & 0\\
0 & 0 & 1\\
\end{pmatrix}
\ee
Performing the product one obtains the expression
\be
\addtolength{\arraycolsep}{3pt}
V_{Hd}= \begin{pmatrix}
c_{12}^d c_{13}^d & s_{12}^d c_{13}^d e^{-i\delta^d_{12}}& s_{13}^d e^{-i\delta^d_{13}}\\
-s_{12}^d c_{23}^d e^{i\delta^d_{12}}-c_{12}^d s_{23}^ds_{13}^d e^{i(\delta^d_{13}-\delta^d_{23})} &
c_{12}^d c_{23}^d-s_{12}^d s_{23}^ds_{13}^d e^{i(\delta^d_{13}-\delta^d_{12}-\delta^d_{23})} &
s_{23}^dc_{13}^d e^{-i\delta^d_{23}}\\
s_{12}^d s_{23}^d e^{i(\delta^d_{12}+\delta^d_{23})}-c_{12}^d c_{23}^ds_{13}^d e^{i\delta^d_{13}} &
-c_{12}^d s_{23}^d e^{i\delta^d_{23}}-s_{12}^d c_{23}^ds_{13}^d e^{i(\delta^d_{13}-\delta^d_{12})} &
c_{23}^dc_{13}^d\\
\end{pmatrix}
\ee

For completeness, we conclude this subsection extending the discussion above to
the lepton sector.
Similarly to the quark sector, the presence in the LHT model of mirror leptons
introduces two new mixing matrices $V_{H \nu}$ and $V_{H\ell}$, in addition to
$V_\text{PMNS}$ \cite{pmns} describing the SM lepton flavour violating interactions.
$V_{H \nu}$ is involved in the interactions of an ordinary neutrino and a
mirror lepton (charged or neutral), while $V_{H\ell}$ appears in the
interactions of an ordinary charged lepton with a mirror lepton (charged or 
neutral).
These $3 \times 3$ unitary matrices satisfy
\be
V_{H \nu}^\dagger V_{H\ell} = V_\text{PMNS}\,,
\label{eq:VHnuVHl}
\ee
with the Majorana phases in $V_\text{PMNS}$ set to zero, as no Majorana masses
have been introduced for the right-handed neutrinos in the LHT model.
The procedure of counting the independent parameters in the mixing matrices is
in the lepton sector the same as in the quark sector.
It follows, then, that $V_\text{PMNS}$ contains $3$ mixing angles and
$1$ phase like $V_\text{CKM}$, while $V_{H \nu}$ is described by $3$ mixing
angles and $3$ phases like $V_{Hd}$. 
Finally, $V_{H\ell}$ can be extracted from~(\ref{eq:VHnuVHl}).  

\subsection{General Counting of Parameters}
The number of independent parameters required to describe the $V_{Hd}$ matrix
can  also be deduced from a more general approach, which allows us to
count the number of physical parameters of a particular sector of a  model already at the level of the basic Lagrangian \cite{counting}.
For simplicity, we will again  consider first the SM and then extend the
discussion to the  LHT model.

In the SM, not all of the 18 moduli and 18 phases of the Yukawa coupling
matrices are physical. In the limit $Y_U = Y_D =0$ the
Lagrangian of the SM has an enlarged chiral symmetry under which the quark
fields transform as \cite{mfvearly}
\be
G_q^\text{SM}=SU(3)_Q \otimes SU(3)_U \otimes SU(3)_D \otimes
U(1)_B  \otimes U(1)_Y  \otimes U(1)_{PQ}\,.
\ee
Using this flavour symmetry
allows us to count the number of physical parameters, like masses,
mixing angles and CP-violating phases hidden in the Yukawa coupling matrices.
A simultaneous transformation of fields and  Yukawa couplings by
\begin{eqnarray}
&u_R \rightarrow V_U u_R, \quad d_R \rightarrow V_D d_R, \quad Q_L \rightarrow V_Q Q_L,\\
&Y_U \rightarrow V_Q Y_U V_U^\dagger, \\
&Y_D \rightarrow V_Q Y_D V_D^\dagger,
\end{eqnarray}
defines an equivalence class of indistinguishable parameterizations.
We can count the physical moduli and phases by $N_ {\rm phys} =
N_{\rm Yukawa} - N_{G} + N_{H}$, where $N_{\rm Yukawa}$,
$N_G$ and $N_{H}$ are the number of moduli and phases of
the Yukawa couplings $Y_U$ and $Y_D$, of the chiral flavour group $G_q$
and of the subgroup $H$ of $G_q$ that leaves $Y_U$ and $Y_D$ invariant,
respectively.\footnote{In the quark sector, $H$
is the baryon number $U(1)_B$. }

The Yukawa couplings $Y_U$ and $Y_D$ are complex $3\times 3$
matrices with 9 moduli and 9 phases each. The flavour group
consists of three $SU(3)$ chiral transformations each parameterized
by 3 moduli and 5 phases and three $U(1)$'s. Hence, we find
\begin{eqnarray}
N_{\rm physical}^{\rm SM} &=& ({\rm moduli, phases})\nonumber\\
&=& (18,18)_{\rm Yukawa} - (9,18)_G + (0,1)_{H}\nonumber\\
&=& (9,1)_{\rm physical}
\end{eqnarray}
corresponding to 6 quark masses, 3 mixing angles and 1 CP-violating
phase of the CKM matrix
in the SM.

Now we are going to apply this method of parameter counting to the LHT
model.\footnote{A detailed description of the LHT model can be found
  e.\,g. in \cite{LHTphen,BBPRTUW}.} In the LHT model, there exist two left-handed $SU(5)$ fermion multiplets
$\Psi^i_1$ and $\Psi^i_2$.
The SM and mirror quarks are contained in the T-invariant linear
combinations of these fields. In addition, there is a right-handed
T-odd $SO(5)$ multiplet $\Psi^i_R$ and the right-handed SM fields
$u^i_R,\;d^i_R$. Note that in the following discussion we neglect
  the presence of the heavy singlet quark fields $T_+$ and $T_-$ for simplicity.

The Yukawa terms for the ordinary up and down quarks are given by \cite{LHTphen,Tobe}
\bea
\mathcal{L}_\text{up} &=&
-\frac{1}{2\sqrt{2}}\lambda^{ij}_u f
  \epsilon_{abc}\epsilon_{xy}\left[(\bar
    \Psi^i_1)_a(\Sigma)_{bx}(\Sigma)_{cy} - (\bar \Psi^i_2 \Sigma_0)_a (\tilde
    \Sigma)_{bx}(\tilde \Sigma)_{cy}\right] u^j_R+h.c.\,,\label{eq:Lup}\\
\mathcal{L}_\text{down}&=&
\frac{i\lambda^{ij}_d}{2\sqrt{2}}f
\epsilon_{ab}\epsilon_{xyz}\left[(\bar \Psi^i_2 )_x (\Sigma)_{ay}
  (\Sigma)_{bz}X-(\bar \Psi^i_1\Sigma_0)_x (\tilde \Sigma)_{ay}(\tilde
 \Sigma)_{bz}\tilde X\right] d^j_R + h.c.\,,
\eea
and the term generating the mirror quark masses reads \cite{Low}
\be\label{eq:Dirac}
\mathcal{L}_\text{mirror}=-\kappa_{ij}f\left(\bar\Psi_2^i\xi +
  \bar\Psi_1^i\Sigma_0\Omega\xi^\dagger\Omega\right)\Psi_R^j+h.c.\,.
\ee
Here $\Sigma_0,\;\Sigma,\;\tilde\Sigma,\;X,\;\tilde X,\;\xi$ and
$\Omega$ are flavour independent quantities, and consequently their
specific form is irrelevant for our counting.  We recall that
$\Psi^i_1$ and $\Psi^i_2$ are related due to T-parity through
$\Psi^i_1 \mapsto -\Sigma_0\Psi^i_2$.
Thus the equality of the coefficients of the two terms in
\eqref{eq:Lup}--\eqref{eq:Dirac} is an immediate consequence of T-parity.

Naively, one would expect that each of the above quark fields transforms under
an independent $U(3)$, thus leading to the following
transformations of the fields and Yukawa couplings
\begin{gather}
\Psi_1 \to V_1 \Psi_1\,,\qquad \Psi_2 \to V_2 \Psi_2\,,\\
\Psi_R \to V_R \Psi_R\,,\qquad u_R \to V_u u_R\,,\qquad d_R \to V_d
d_R\,,\\
\kappa \to V_1 \kappa V_R^\dagger\,,\qquad \kappa \to V_2 \kappa
V_R^\dagger\,,\label{eq:kappa}\\
\lambda_u \to V_1 \lambda_u V_u^\dagger\,,\qquad \lambda_u \to V_2
\lambda_u V_u^\dagger\,,\\
\lambda_d \to V_1 \lambda_d V_d^\dagger\,,\qquad \lambda_d \to V_2
\lambda_d V_d^\dagger\,.\label{eq:lambdad}
\end{gather}
As a consequence of T parity, however, we see that in order to leave   \eqref{eq:Lup}--\eqref{eq:Dirac}
invariant we have to choose
\be
V_1\equiv V_2
\ee
in the transformations \eqref{eq:kappa}--\eqref{eq:lambdad}.

Thus we now find the flavour symmetry group
\be
G_q^\text{LHT} = U(3)_1 \otimes U(3)_R \otimes U(3)_u \otimes U(3)_d\,,  
\ee
which can be rewritten as follows
  \begin{equation}
G_q^\text{LHT} = SU(3)_Q \otimes SU(3)_U \otimes SU(3)_D \otimes
U(1)_B \otimes U(1)_{PQ}\otimes U(1)_{Y_1} \otimes SU(3)_{R}  \otimes U(1)_{Y_2}.
\end{equation}

Since $\kappa$, $\lambda_u$ and $\lambda_d$ are
complex $3\times 3$ matrices, they have each in principle 9 moduli and 9 phases.
Due to the flavour symmetry group $G_q^\text{LHT}$ one
can remove 12 moduli and $24-1$ phases, i.\,e. counting the physical moduli
and phases now yields
\begin{eqnarray}
N_{\rm physical}^{\rm LHT} &=& ({\rm moduli, phases})\\
&=& (27,27)_{\rm Yukawa} - (12,24)_{G}+(0,1)_H\\
&=& (15,4)_{\rm physical}
\end{eqnarray}
corresponding to 3 additional masses, 3 additional mixing angles and 3
new phases in addition to the 6 SM masses, 3 CKM mixing angles and 1
CKM phase. We note that we only count three equal masses,
corresponding to three mirror fermion generations, since at first order in
the $v^2/f^2$ expansion up and
down mirror fermions are degenerate in mass. This degeneracy is broken
when higher $v^2/f^2$ corrections are taken into account.

{ Including now also $T_+$ and $T_-$ to the Lagrangian does not
  affect the above counting, but merely introduces a single new
  parameter $x_L$ wich parameterizes both the masses of $T_+,T_-$ and
  the mixing of $T_+$ with the standard top quark.}

\newsection{Conclusions}
In this note we have demonstrated that in contrast to the CKM matrix 
and to what was claimed in the analyses in \cite{Hubisz} and in the first
version of \cite{BBPTUW}, the mixing matrix $V_{Hd}$ of the LHT
model contains, in addition to three mixing angles, also 
three physical complex phases $\delta_{12}^d$, $\delta_{13}^d$ and 
$\delta_{23}^d$. The analyses presented in \cite{Hubisz,BBPTUW} apply only to 
situations in which $\delta_{12}^d$ and $\delta_{23}^d$ are set to zero.
{This assumption is quite reasonable, since the impact of the additional
  two phases is numerically small, once all existing constraints 
on FCNC processes are 
taken simultaneously into account, and does not change qualitatively the
  results of \cite{BBPTUW}.}
 The analysis in \cite{BBPRTUW} presents in 
addition to scenarios with $\delta^d_{12}=\delta^d_{23}=0$ a global
analysis of FCNC processes in which all phases are treated as free
parameters. 

The main message of the new version of \cite{BBPTUW} and of the 
rare decay analysis in \cite{BBPRTUW} is then the following one. Even for
 $\delta^d_{12}=\delta^d_{23}=0$ large deviations from the SM
expectations, in particular for rare $K$ decays and the CP asymmetry 
$S_{\psi\phi}$, are possible. 
The inclusion
of two new phases does not change this picture qualitatively, although at
the quantitative level spectacular effects in certain observables are
easier to obtain.

\subsection*{Acknowledgements}

This research was partially supported by the German `Bundesministerium f\"ur 
Bildung und Forschung' under contracts 05HT4WOA/3, 05HT6WOA.

\end{document}